\documentclass[aps,reprint,showpacs,superscriptaddress]{revtex4-1}
\usepackage{graphicx}
\usepackage{dcolumn}   % needed for some tables
\usepackage{bm}        % for math
\usepackage[dvipsnames]{xcolor}
\usepackage{cancel}
\usepackage{amsfonts}
\usepackage{amssymb, amsmath,framed,mathtools}
\usepackage{hyperref}
\hypersetup{
    bookmarks=true,         % show bookmarks bar?
    unicode=false,          % non-Latin characters in Acrobatâs bookmarks
    pdftoolbar=true,        % show Acrobatâs toolbar?
    pdfmenubar=true,        % show Acrobatâs menu?
    pdffitwindow=false,     % window fit to page when opened
    pdfstartview={FitH},    % fits the width of the page to the window
    pdfsubject={Plasma fluid theory},   % subject of the document
    pdfproducer={GNU Make}, % producer of the document
    pdfkeywords={keyword1} {key2} {key3}, % list of keywords
    pdfnewwindow=true,      % links in new PDF window
    colorlinks=true,        % false: boxed links; true: colored links
    linkcolor=blue,      % color of internal links (change box color with linkbordercolor)
    citecolor=blue,         % color of links to bibliography
    filecolor=blue,      % color of file links
    urlcolor=blue           % color of external links
}
\usepackage{color}

\usepackage{bm,suffix}
\expandafter\ifx\csname package@font\endcsname\relax\else
 \expandafter\expandafter
 \expandafter\usepackage
 \expandafter\expandafter
 \expandafter{\csname package@font\endcsname}
\fi
 \DeclarePairedDelimiterX\MeijerM[3]{\lparen}{\rparen}%
{\begin{smallmatrix}#1 \\ #2\end{smallmatrix}\delimsize\vert\,#3}

\newcommand\MeijerG[8][]{%
  G^{\,#2,#3}_{#4,#5}\MeijerM[#1]{#6}{#7}{#8}}

\WithSuffix\newcommand\MeijerG*[7]{%
  G^{\,#1,#2}_{#3,#4}\MeijerM*{#5}{#6}{#7}}

\def\bq{\begin{equation}}
\def\eq{\end{equation}}
\def\bqy{\begin{eqnarray}}
\def\eqy{\end{eqnarray}}
\def\calc{\mathcal{C}}

\def\cale{\mathcal{E}}
\def\calf{\mathcal{F}}

\def\calh{\mathcal{H}}
\def\cali{\mathcal{I}}

\begin{document}

\title[Distribution of 3D Plasmoids]{A Maximum Entropy Principle for inferring the Distribution of 3D Plasmoids}

\author{Manasvi Lingam}
\email{manasvi.lingam@cfa.harvard.edu}
\affiliation{Harvard-Smithsonian Center for Astrophysics, Cambridge, MA 02138, USA}
\affiliation{John A. Paulson School of Engineering and Applied Sciences, Harvard University, Cambridge, MA 02138, USA}
\author{Luca Comisso}
\email{lcomisso@princeton.edu}
\affiliation{Department of Astrophysical Sciences, Princeton University, Princeton, NJ 08544, USA}
\affiliation{Princeton Plasma Physics Laboratory, Princeton, New Jersey 08543, USA}

\begin{abstract}
The Principle of Maximum Entropy, a powerful and general method for inferring the distribution function given a set of constraints, is applied to deduce the overall distribution of 3D plasmoids (flux ropes/tubes) for systems where resistive MHD is applicable and large numbers of plasmoids are produced. The analysis is undertaken for the 3D case, with mass, total flux and velocity serving as the variables of interest, on account of their physical and observational relevance. The distribution functions for the mass, width, total flux and helicity exhibit a power-law behavior with exponents of $-4/3$, $-2$, $-3$ and $-2$ respectively for small values, whilst all of them display an exponential falloff for large values. In contrast, the velocity distribution, as a function of $v = |{\bf v}|$, is shown to be flat for $v \rightarrow 0$, and becomes a power law with an exponent of $-7/3$ for $v \rightarrow \infty$. Most of these results are nearly independent of the free parameters involved in this specific problem. A preliminary comparison of our results with the observational evidence is presented, and some of the ensuing space and astrophysical implications are briefly discussed.
\end{abstract}

\maketitle

\section{Introduction} \label{SecIntro}
It is now widely acknowledged that magnetic reconnection, which entails the conversion of magnetic energy into charged particle energy, drives many astrophysical phenomena in the Universe. Some well-known examples include magnetospheric substorms, gamma ray bursts, stellar and solar flares \citep{TS97,Kuls05,BP07,ZwYa09,JD11,KZ15}. Under normal circumstances, this fast conversion would not be possible for highly elongated current sheets, as the conventional estimates for the reconnection rate are too low. Fortunately, it has been known since the 1980s that such current sheets are subject to a violent instability that was dubbed the plasmoid instability later \citep{BHYR}, as it results in the formation of plasmoids \citep{Bisk86,Bisk00}.

The plasmoid instability breaks up the elongated current sheets, and leads to very high reconnection rates that are nearly independent of the Lundquist number $S$ in the nonlinear regime \citep{BHYR,CSD09,DRAKYB,HB10,LSSU12,HB13,CGW15}. The emergence of this fast reconnection process, mediated by the plasmoid instability, has been shown to have several important consequences. Plasmoid reconnection can play an important role in regulating solar flares \citep{KKB00,ShiTa01}, coronal mass ejections \citep{LMS15}, chromospheric jets \citep{NKLW15}, particle acceleration \citep{Oi02,DSCS06,Chet08,Birn12}, pulsar winds \citep{SirSpi14,GLDL14}, blazar emissions and jets \citep{SPG15,PGS16}, as well as a wide range of laboratory and fusion experiments \citep{EBra15,JAJYYF,EBra16}.

In plasmoid-mediated reconnection, a great deal of analytic and numerical work has been undertaken for the linear regime. In Sweet-Parker current sheets, scaling relations for the growth rate and number of plasmoids produced were predicted by \citet{TS97} and \citet{LSC07} as a function of $S$, and were generalized to encompass arbitrary magnetic Prandtl numbers in \citet{ComGra16}. However, in dynamically evolving current sheets, the final Sweet-Parker aspect ratio cannot be achieved, as they disrupt prior to this stage \citep{Bisk82,CLHB16,Ten16,CLHB17,HCB17}. Recently, \citet{CLHB16,CLHB17} utilized a principle of least time to formulate a general theory for the onset and development of the linear phase of the plasmoid instability. They demonstrated that the scalings were no longer true power laws, and depended on other parameters, namely the initial noise inherent in the system, the characteristic rate of current sheet evolution, and the nature of the thinning process.

From the observational standpoint, it is not very feasible to verify the linear stage of the plasmoid instability. As a result, a great deal of interest has been centred on the nonlinear phase of this instability. One of the most notable amongst them is the distribution of plasmoids as a function of their intrinsic properties, such as their flux and width. From the theoretical standpoint, there have been three major works in this subject, albeit only for plasmoids emergent from a 2D reconnection process. \citet{ULS10} and \citet{HB12} proposed phenomenological kinetic models for the flux distribution function, and arrived at scaling laws with exponents of $-2$ and $-1$ respectively. On the other hand, \citet{FDS10} prescribed rules for the generation, growth and merging of plasmoids, and arrived at an integro-differential equation for the distribution function that depended on the flux and the area.

There are, arguably, two important factors that need to be taken into account when paving the way for a more complete theory. Firstly, the distribution function is obviously statistical, as a result of which it would clearly be ideal to rely on a more statistical-mechanical approach based on first principles. Secondly, many of the theoretical and numerical studies in the area have relied on heuristic models with simplified geometries to arrive at their results. Hence, except in certain circumstances, the resultant parameters and the distribution function may not constitute true observables.

It is, therefore, the goal of this work to construct a 3D plasmoid distribution function in the highly nonlinear regime, characterized by a large number of plasmoids, that depends on three genuinely physical observables, namely the plasmoid mass, total flux and velocity. During the course of the paper, we shall use the nomenclature ``plasmoids'' to describe magnetic islands in three dimensions, i.e. flux ropes/tubes \citep{MH91}. In 3D geometry, the magnetic field lines could become stochastic, and the flux surfaces may not be well-defined \citep{BGPCPF,HuBha16}. Yet, one can still identify coherent structures which can be treated on the same footing as flux ropes. Hence, in this paper, we shall operate with the implicit assumption that well-defined coherent structures (or flux ropes) are existent. 

For the purpose of this work, we adopt the standard resistive magnetohydrodynamics (resistive MHD) as the underlying physical model. Our results are therefore relatively valid for astrophysical plasmas where resistive MHD constitutes a reasonable description of the plasma dynamics. Hence, for plasma environments such as planetary magnetospheres \citep{BMT16} where kinetic effects become important \citep{JD11}, our results are not likely to be accurate. In contrast, resistive MHD constitutes a useful model for certain astrophysical systems like solar and stellar coronae \citep{Pri14} and the interstellar medium \citep{Zwei99}, and therefore we anticipate that our findings have a higher degree of applicability to these systems.

Our paper is centered around the Principle of Maximum Entropy \citep{Jay03}, which has been successfully used in many fields of science. We shall derive the overall 3D distribution function, examine the various asymptotic limits, and eventually discuss its potential connections with astrophysics. The outline of the paper is as follows. A short introduction to the Principle of Maximum Entropy is offered in Sec. \ref{SecMaxEnt}. The physical parameters of interest, and the relevant constraints are introduced in Sec. \ref{SecDerPDF}, and the overall plasmoid distribution function is derived. In Sec. \ref{SecAnDF}, the single-variable distribution functions for the mass, width, total flux, helicity and velocity are obtained. Sec. \ref{SecAstroImp} contains a discussion of the relevance of our work in space and astrophysical plasmas, and we finally summarize our results in Sec. \ref{SecConc}.

\section{On the Principle of Maximum Entropy}\label{SecMaxEnt}
In this Section, we shall briefly elucidate some of the salient features of the Principle of Maximum Entropy (MaxEnt), which has been used quite extensively in many areas of astronomy and astrophysics \citep{LB67,NaNi86,BT87}. We also discuss its relevance in the context of determining the statistical distribution of 3D plasmoids.

\subsection{What is MaxEnt and where is it used?}\label{SSecMEInt}
In equilibrium statistical mechanics, it is known that the micro-, macro- and grand canonical distribution functions can be derived through a common principle. The central idea relies upon the maximization of the entropy $S = -\sum_i p_i \ln p_i$ with respect to the probabilities $p_i$, subject to holding the appropriate constraints on the system fixed \citep{Tol38,Haar55,Balescu75,Call85}. By drawing upon the classical ideas of Boltzmann, Gibbs and Shannon, E.T. Jaynes explored the connections between information theory and statistical mechanics and argued that statistical mechanics could be interpreted as a method for inferring the probability distributions based on the (limited) data available \citep{Jay57a,Jay57b,Jay03}. 

It is instructive to consider Jaynes' own words \citep{Jay57a} in this regard:
\emph{``In the resulting ``subjective statistical mechanics,'' the usual rules are thus justified independently of any physical argument, and in particular independently of experimental verification; whether or not the results agree with experiment, they still represent the best estimates that could have been made on the basis of the information available.''} 

MaxEnt has enjoyed a considerable degree of success ranging from fields as disparate as ecology \citep{Harte11} and climate science \citep{Pal79,OOLP03} to gravitational dynamics \citep{LB67,Pad90,SpHe92,Naka00,HTR17} and plasma physics \citep{HB87,YM08,DHMMH08}. We shall not go into further details of MaxEnt as there exist several excellent reviews and summaries of the subject \citep{SJ80,Gra80,Jay82,Jay03,Dew03,Dew05,MS06,PGLD13}, and detailed analyses of the arguments for and against MaxEnt can also be found therein.
%In some of these works, a Minimum Entropy Principle (MinEnt) was utilized as opposed to the standard MaxEnt. We refer the reader to \citet{Jay80} and \citet{Lucia13} for a discussion of the subtle connections between these two, outwardly conflicting, theories.

\subsection{MaxEnt and the distribution of plasmoids}
In dealing with the plasmoid distribution function, it is clear that the system cannot necessarily be regarded as being in equilibrium throughout, that the plasmoids themselves are complex entities (unlike, for e.g., ideal gas molecules), etc. Fortunately, as per the discussion in Sec. \ref{SSecMEInt}, many of these points are rendered moot. 

Firstly, we recall that MaxEnt must be understood as a fairly general statement regarding statistical inference \footnote{This fact served as one of the primary motivations behind our use of the word ``infer'' in the title.}, and its generalization, Maximum Caliber (MaxCal), is valid when dealing with non-equilibrium systems \citep{Jay80,PGLD13,HWDD15}. Since MaxEnt is applicable to complex systems close to equilibrium \citep{Haken06}, it could potentially be used for deriving the plasmoid distribution function. 

A comment regarding the choice of the entropy functional is in order. At times, there is a tendency to associate the Boltzmann-Gibbs-Shannon (BGS) entropy \citep{Jay65} only with the presence of an associated kinetic equation \footnote{Although we shall use the nomenclature BGS entropy henceforth, it is important to recognize that there exist subtle differences between the Boltzmann, Gibbs and Shannon formulations \citep{Jay65}.}. In actuality, the BGS entropy has been ``postulated'' in a more general and axiomatic manner \citep{Penro79,Call85}. Moreover, the BGS entropy is endowed with the necessary mathematical and physical requirements, such as invariance, concavity, and additivity to name a few \citep{Weh78,Mack89}. 

We would like to mention some important caveats at this stage. Since we are dealing with a statistical treatment, there is an implicit assumption that there exist a sufficiently high number of plasmoids that interact amongst each other; as the theory is ``thermodynamic'' in a certain sense, a large number of interacting entities is ostensibly necessary. In the magnetospheres of planets (and satellites), plasmoids are produced in low numbers, as noted in Sec. \ref{SecAstroImp}, implying that the MaxEnt formulation may not yield accurate results. On the other hand, since a large number of plasmoids are produced in the solar corona \citep{Jan16}, and given the applicability of resistive MHD in this regime, it could be more reasonable to employ MaxEnt in this context. The presence of a large number of plasmoids is also important from the standpoint of observations (or simulations), as otherwise proper statistics cannot be deduced. 

Hence, bearing the above properties and caveats in mind, we are now in a position to adopt the standard approach of \citet{Jay57a} and construct a variational principle wherein the BGS entropy functional can be extremized whilst holding the necessary constraints fixed. 

\section{The derivation of the Plasmoid Distribution Function} \label{SecDerPDF}
Here, we shall delineate some general properties of the 3D plasmoid distribution function $\mathcal{F}$, motivate the appropriate constraints, set up the variational principle, and thereby arrive at the plasmoid distribution function. 

\subsection{Properties of the plasmoid distribution function}
Let us begin by considering the variables that the plasmoid distribution function $\calf$ depends upon. We begin by noting that our ``system'' is now taken to comprise of a sufficiently large collection of plasmoids, and not the entire current sheet as a whole.

There are many properties that can characterize a particular plasmoid. These include:
\begin{itemize}
\item The mass $m$ enclosed within a given plasmoid.
\item The density $\rho$ of a particular plasmoid. Notice that a knowledge of $m$ and $\rho$ suffices to determine the volume $V$ of the plasmoid. 
\item The total flux $\psi$ enclosed in the plasmoid. The use of the word ``total'' is deliberate, as we are dealing with the 3D setup. It also constitutes an observationally relevant physical variable, particularly in the solar context \citep{WLS05,Demo08,VS10}.
\item The velocity ${\bf v}$ associated with the motion of the plasmoid. 
\item The position ${\bf x}$ describing the spatial location of the plasmoid.
\end{itemize}
In addition to the above quantities, the plasmoid distribution function can be expected to evolve over time as well. $\calf$ could also depend on other properties, but we believe that we have listed all of the most salient candidates. Clearly, retaining all of these variables would make our procedure very onerous and rob it of physical transparency. Hence, we shall introduce the following simplifications:
\begin{itemize}
\item We are ultimately interested, by means of MaxEnt, to determine the final $\calf$ that the system relaxes to. In this case, an explicit dependence on time is not expected to be present. 
\item We shall assume that the plasma inside the plasmoids is incompressible, i.e. we shall take $\rho = \mathrm{const}$, and hence it does not enter our model.
\item Handling a spatial dependence is more difficult, and we shall consider the cases where $\calf$ does not depend on ${\bf x}$, which may be valid when the system has a weak spatial dependence \footnote{Alternatively, perhaps, one could interpret $\calf$ as the coarse-grained distribution function where the ${\bf x}$ dependence has already been eliminated.}.
\end{itemize}
Thus, based on the above set of postulates, the distribution function has the form $\calf\left({\bf v},\psi,m\right)$. As with the previous theories, the distribution function $\calf$ and its constituent variables are continuous.

Although we know that $m \propto V$, we still do not know how the volume depends on the half-width $R$ and on the area. To resolve this matter, we must introduce an additional simplification. In 3D, it is natural to speak of flux ropes or flux tubes \citep{RPL90,BP07}, and our choice of the word ``plasmoids'' should actually be taken to refer to such entities. These structures can be envisioned as cylinders with radius $R$ and height $H$, and we shall introduce the assumption $\lambda = H/R = \mathrm{const}$. Our ansatz is supported by recent numerical simulations undertaken by \citet{HuBha16} with a finite guide field, wherein cylindrical-shaped flux ropes were approximately characterized by a scale-independent anisotropy.

This does not significantly reduce the scope of our analysis, since one can easily make $\lambda$ a function of $R$. The current simplification amounts to the statement that all plasmoids have a self-similar structure, i.e. their volumes and areas may be different, but they have the same value of height to radius. Such ansatzen have already been used quite often in other areas of physics, where the basic ``building blocks'' are extended objects, such as liquid crystals and polymers \citep{RFL59,SS74,VL92}. We finally end up with
\begin{equation} \label{mRrel}
\frac{m}{\rho} = V = \pi R^2 H = \lambda \pi R^3
\end{equation}

\subsection{The choice of constraints} \label{SSecCons}
Before exploring the constraints that need to be included in the variational principle, a brief discussion of the dynamics is necessary \citep{BVK08}. There are three primary processes involved:
\begin{itemize}
\item \underline{Process I:} New plasmoids are ``born'', and they ``grow'' by accreting mass and flux, growing bigger in size.
\item \underline{Process II:} Two plasmoids can ``collide'' and thereby ``merge''. Some of the merging rules have already been explored (in the 2D setting) by \citet{FDS10} and \citet{HB12}. 
\item \underline{Process III:} Plasmoids can be effectively said to have ``died'' when they have been ejected out of the current sheet. 
\end{itemize}

Let us now take a look at some of the potential constraints that are feasible. \\

{\bf Total number:} Although Processes I and III create and destroy plasmoids, there is no guarantee that they balance each other. In addition, Process II does not conserve the number of plasmoids, since the total reduces by unity. Therefore, taken collectively, there is no reason to believe that the total number $N = \int_\Omega \calf\,dm\,d\psi\,d{\bf v}$ is conserved. Even if this constraint were incorporated, the overall effect would be qualitatively unimportant since the distribution function would just end up being multiplied by a constant factor.\\

{\bf Total mass:} It is clear that the mass must be conserved during Process II. However, there are two other processes, one of which depletes the mass and the other that replenishes it. If we assume that, even though each of the processes are dynamical, the whole system can exist in a state of detailed balance, enabling the total mass to be conserved. Thus, the total mass is
\begin{equation} \label{Mass}
M = \int_\Omega \calf m\,dm\,d\psi\,d{\bf v},
\end{equation}
where $\Omega$ is the volume of the total phase space.\\

{\bf Total momentum:} It is tempting to argue that a similar set of arguments should hold true for momentum conservation. In the 2D case, at least, one does \emph{not} expect the total momentum (along each direction) to be conserved, owing to the fact that there exist clear inflow and outflow directions. Some processes can occur only along the former, and others along the latter alone. Thus, there is a fundamental asymmetry with respect to the axes for all the three processes, implying that momentum conservation along all the directions ought not be possible.\\

{\bf Total flux:} When we consider flux, it is known that flux is \emph{not} conserved during the merging of two plasmoids, viz. in Process II \citep{HB12}. To compensate for this absence of conservation, Processes I and II would need to self-adjust in a very precise manner for flux conservation to hold true. If this condition were indeed valid, the total flux
\begin{equation} \label{Flux}
\Psi =  \int_\Omega \calf \psi\,dm\,d\psi\,d{\bf v}
\end{equation}
would be conserved. In (\ref{Flux}), the distinction between the total flux $\Psi$ of the complete system and the total flux $\psi$ of a given plasmoid must be duly borne in mind; the latter has the standard definition, $\psi = \int {\bf B}\cdot d{\bf S}$. \\

{\bf Total energy:} It is well known that the conservation of total energy is closely linked with time-translation symmetry \citep{LanLi60,Arnold89}. Taken individually, none of the processes can conserve the energy, but together, there is a possibility for doing so. If energy conservation is indeed valid for our system, then
\begin{equation} \label{Energy}
E = \frac{1}{2} \int_\Omega \calf \left[m v^2 + \calc \psi^2 \left(\frac{m}{\rho}\right)^{-1/3}\right]\,dm\,d\psi\,d{\bf v},
\end{equation}
and the first term is clearly the kinetic energy, whilst the second term is the magnetic energy that is discussed further below. An important point worth noting here is that we have used the MHD energy, since we are interested in systems where collisionless effects are not important. For instance, if we consider a domain where extended MHD is applicable, a third term must be included in the above formula that will be proportional to the square of the electron skin depth in normalized units \citep{Lu59,KLM,CA14,LMM15}. Similarly, when Finite Larmor Radius (FLR) are important, they can be incorporated into the above formula by adding gyroaveraged contributions \citep{SH01,Scott,CGWB}. We have neglected thermal energy in our analysis, as this introduces a further level of complexity by means of an extra variable. Moreover, many plasmas are characterized by $\beta \ll 1$ \citep{GP04,Kuls05}, which is consistent with the neglect of the thermal energy. Furthermore, even with the introduction of a finite thermal energy that is proportional to the temperature, we have verified that the overall scalings remain unaltered. 

In the energy, $\calc$ is a dimensionless constant that depends upon $\lambda$ and other numerical factors (such as $\pi$). We are not interested in the exact expression for $\calc$ as we seek instead to evaluate the \emph{algebraic} dependence of $\calf$. We have also assumed that our system has a weak guide field. Our procedure can also be generalized to include an arbitrary guide field, and will result in an extra term that is linearly proportional to $m$ in (\ref{Energy}). It can therefore be easily shown that the scaling relations for the mass (also the volume and radius) distribution function derived in Sec. \ref{SSecMWDF} for the low-mass limit remain unchanged, as well as the overall behavior. Moreover, the flux and velocity distributions in both asymptotic regimes are also unaffected. Hence, the comparison of our theoretical results with the observational data undertaken in Sec. \ref{SecAstroImp} is valid even when a strong guide field is present.

It is easy to verify that the second term represents the magnetic energy by carrying out a simple dimensional analysis. A second method for arriving at this expression is via the magnetic helicity. Suppose that the magnetic energy and helicity of a plasmoid are represented by $\cale_B$ and $\calh$ respectively. From dimensional considerations, it is once again evident that
\begin{equation} \label{MagEHelRel}
\cale_B \sim \ell^{-1} \calh \propto \left(\frac{m}{\rho}\right)^{-1/3} \calh,
\end{equation}
where the second equality follows by observing that $\ell$, the characteristic length scale, must correspond to either $R$ or $H$. Either of these two quantities can be determined in terms of $m$ via (\ref{mRrel}). Thus, if we show that $\calh \propto \psi^2$, we will recover the magnetic energy in (\ref{Energy}). This is explored below, when the total helicity is introduced as the next constraint. 

The first equality in (\ref{MagEHelRel}) can also be justified on more rigorous grounds than dimensional analysis. We have already stated that our plasmoids are now modelled as flux tubes. For some special configurations, wherein the sum of the twist and the writhe is zero, it is known that
\begin{equation}
\cale_B \geq \alpha \calh,
\end{equation}
where $\alpha \propto V^{-1/3}$ \citep{Ric08}, and the $m$-dependence can be determined via (\ref{mRrel}). The above inequality, which also goes by the name of the `realizability condition', was first proven by Arnold in the 1970s \citep{Arnold89,AK98}.

The above approach relied upon topological fluid mechanics, and the reader may consult \citet{LMM16} for a recent overview of this subject for a diverse array of plasma fluid models. There is also a more direct means of arriving at $\cale_B$ in (\ref{Energy}) via this area of research. It was shown in \citet{Moff90} that the minimum value of the magnetic energy attained through a relaxation process is $\mu\, \psi^2 V^{-1/3}$, where $\mu$ is an invariant that solely depends on the magnetic topology and can be computed exactly in certain scenarios \citep{MoCh95}. Clearly, upon using (\ref{mRrel}), this expression has the same form as the magnetic energy delineated in (\ref{Energy}). \\

{\bf Total helicity:} Most of the interest in the total magnetic helicity stems from its unique topological properties \citep{BF84,Berg99}. There is considerable experimental and theoretical evidence that supports the robustness of helicity conservation even in the presence of magnetic reconnection \citep{Berg84,PfGe91,KI13,SKPKI}. Hence, it is quite plausible that the total magnetic helicity of the system is also conserved. Then, we end up with
\begin{equation} \label{Helicity}
K = \frac{1}{2} \int_\Omega \calf \psi^2\,dm\,d\psi\,d{\bf v},
\end{equation}
and the factor of $\psi^2$ could be justified, as earlier, through dimensional analysis. It can also be obtained on mathematical grounds - it was shown by \citet{MoRi92} that the helicity, modulo topological invariants (such as the Gauss and Calugareanu-White linking numbers), was quadratic in the flux; we also refer the reader to \citet{Moff90,MoCh95,Berg99,Ric08}. The relation $\calh \propto \psi^2$ is also backed by a fairly high degree of empirical evidence from solar observational data \citep{Demo07,Demo08,BHPvS}. In reality, (\ref{Helicity}) would also involve an extra numerical factor, which can be easily eliminated by simply rescaling the definition of $K$.

\subsection{The formulation of the MaxEnt variational principle}
The BGS entropy functional for our model corresponds to
\begin{equation}
S = - \int_\Omega \calf \ln \calf\,dm\,d\psi\,d{\bf v},
\end{equation}
with $\calf$ denoting the plasmoid distribution function. We are now in a position to write down our variational principle as all the relevant pieces have been assembled. Let us commence our analysis with an important observation. Of all the constraints discussed in Sec. \ref{SSecCons}, only the energy exhibits a complex dependence on \emph{all} three variables, namely $m$, $\psi$ and ${\bf v}$. In contrast, all of the other functionals are dependent on just one variable.

To summarize, the mass (\ref{Mass}), flux (\ref{Flux}), energy (\ref{Energy}) and helicity (\ref{Helicity}) have been chosen as our constraints.  Thus, our variational principle is given by
\begin{equation}
\delta A = 0, \quad A:= S - \beta E - \gamma M - \delta \Psi - \varepsilon K.
\end{equation}
We have chosen the notation $A$ and $\beta$ for the target functional and the Lagrange multiplier preceding the energy. Our choices are deliberate since these quantities can be rightfully seen as the generalizations of the Helmholtz free energy and the inverse temperature (from thermodynamics) respectively \citep{Tol38,Call85}.

Upon variation, we arrive at
\begin{equation} \label{EqDF}
\calf = \exp\left(-\beta \cale - \gamma m - \delta \psi - \frac{\varepsilon \psi^2}{2} - 1\right),
\end{equation}
and if we take $\{\gamma,\delta,\varepsilon\} = 0$, the similarity with the canonical ensemble distribution function \citep{Balescu75} is not very surprising, given that only the energy constraint has been included. Here, it must be noted that
\begin{equation}
\cale = \frac{m v^2}{2} + \frac{\calc}{2} \psi^2 \left(\frac{m}{\rho}\right)^{-1/3},
\end{equation}
is the total plasmoid energy in the MHD regime.

\section{Analysis of the plasmoid distribution function}\label{SecAnDF}
We are now in a position to start analyzing the properties of the plasmoid distribution function (\ref{EqDF}). We shall present the derivation of the distribution functions for the mass, width, total flux, helicity and (3D) velocity in this Section. 

\subsection{The mass and width distributions} \label{SSecMWDF}
We commence our derivation by introducing 
\begin{equation} \label{mpsiDF}
    \calf_{m,\psi} = \int \calf\,d{\bf v},
\end{equation}
which represents the joint distribution function of the plasmoids as a function of $\psi$ and $m$. An inspection of (\ref{EqDF}) reveals that it is dependent \emph{only} on the modulus of ${\bf v}$. Hence, we are free to use the transformation $d{\bf v} = v^2\, \sin \theta\, dv\,d\theta\, d\phi$. In Sec. \ref{SSecCons}, we stated that we shall consider a system where the guide field is absent, implying that one cannot define motion parallel and perpendicular to the guide field. Hence, it seems reasonable to suppose that the conventional ranges for $\theta$ and $\phi$, namely $\left[0,\pi\right]$ and $\left[0,2\pi\right]$ respectively, are valid. This leads us to $d{\bf v} = 4\pi v^2\,dv$, and substituting it into (\ref{mpsiDF}) results in
\begin{eqnarray} \label{mpsiDFfinal}
\calf_{m,\psi} &=& \left(\frac{2\pi}{\beta m}\right)^{3/2} \exp\Bigg[-\frac{\beta \calc}{2}\left(\frac{\rho}{m}\right)^{1/3} \psi^2 \nonumber \\ 
&& \hspace{0.6 in} - \frac{1}{2}\left(2\delta + \varepsilon \psi\right)\psi - \gamma m - 1 \Bigg],
\end{eqnarray}
where the limits of integration are from $v=0$ to $v=\infty$. Next, one can analytically calculate the mass distribution as follows
\begin{equation}
\calf_m = \int_{\psi=0}^{\psi=\infty} \calf_{m,\psi}\,d\psi,
\end{equation}
and the integration limits are taken to be positive as per the standard convention. The final answer is given by
\begin{eqnarray} \label{mDF}
\calf_m &=& \frac{2\pi^2}{\sqrt{\sigma}\left(\beta m\right)^{3/2}} \left[1-\mathrm{Erf}\left(\frac{\delta}{\sqrt{2\sigma}}\right)\right] \nonumber \\
&& \quad \times \exp\left(\frac{\delta^2}{2\sigma}-\gamma m - 1\right),
\end{eqnarray}
where we have introduced the auxiliary function
\begin{equation} \label{defsigma}
\sigma = \varepsilon + \beta \calc \left(\frac{\rho}{m}\right)^{1/3}.
\end{equation}

\begin{figure}
  \centering
  \includegraphics[width=1.0\linewidth]{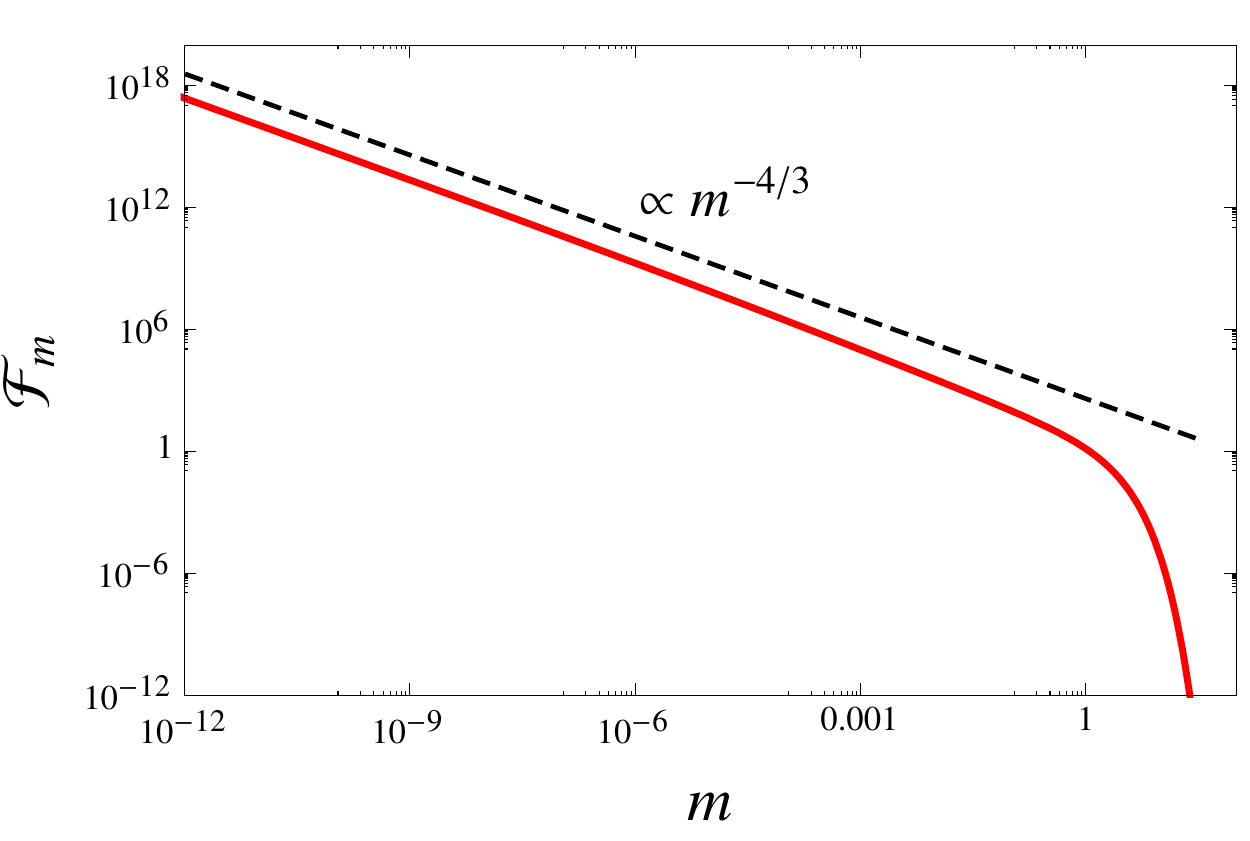}
  \caption{The mass distribution function (\ref{mDF}) is plotted as a function of $m$. The solid red curve represents the exact solution, whilst the dashed black line denotes the asymptotic solution (\ref{m0DF}) that is valid for small $m$. Here, we have used the fiducial values $\beta = \gamma = \delta = \varepsilon = 1$ and $\calc = 1/4\pi$. The latter value was chosen because this factor appears in the magnetic energy.}
  \label{fig:distribution_m}
\end{figure}

We are now in a position to extract the two limits of the mass distribution. We find that
\begin{equation} \label{m0DF}
\lim_{m \rightarrow 0}\, \calf_m \propto m^{-4/3},
\end{equation}
but it is necessary to recognize that the immediate higher-order contribution is $m^{-7/6}$, which is smaller than the leading-order contribution (\ref{m0DF}) only by a factor of $m^{1/6}$ which exhibits a weak dependence on $m$. Hence, the $-4/3$ exponent will not prevail for long, as it will be subsumed by the higher-order contributions at a fairly early stage. In the other limit, we have
\begin{equation} \label{mInfDF}
\lim_{m \rightarrow \infty}\, \calf_m \propto m^{-3/2} e^{-\gamma m},
\end{equation}
and the exponential cutoff is clearly dominant in this region, subsuming the power-law behavior. We have plotted (\ref{mDF}) in Fig. \ref{fig:distribution_m} - the power-law limit for small $m$, namely (\ref{m0DF}), has also been included. It is clear that, for larger values of $m$, the exponential falloff becomes more important. The overall structure of Fig. \ref{fig:distribution_m} is quite similar to the 2D numerical results of \citet{LEKG16} (see their Fig. 10), although our (3D) power-law slope is slightly steeper.

\begin{figure}
  \centering
  \includegraphics[width=1.0\linewidth]{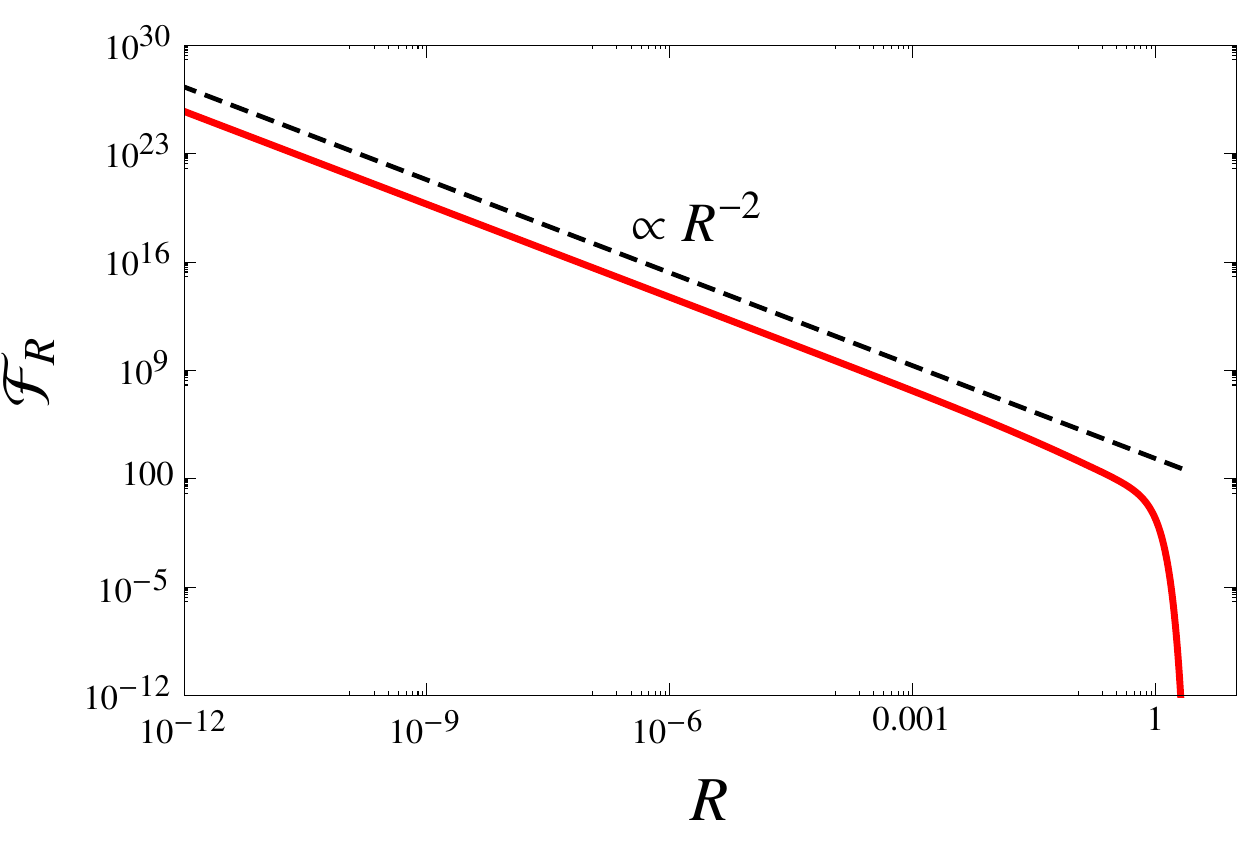}
  \caption{The width distribution function (\ref{RDFfin}) is plotted as a function of $R$. The solid red curve represents the exact solution, whilst the dashed black line denotes the asymptotic solution (\ref{R0DF}), which works well for small $m$. Here, we have used the fiducial values $\beta = \gamma = \delta = \varepsilon = 1$ and $\calc = 1/4\pi$. The latter value was chosen because this factor appears in the magnetic energy.}
  \label{fig:distribution_R}
\end{figure}

Under a change of variables, the probability contained in a differential area is known to be constant \citep{Kolm50}. Let us denote the width distribution of the plasmoids by $\calf_R$. Using this principle, we find
\begin{equation}
\calf_m\, dm = \calf_R\, dR,
\end{equation}
and we can now make use of (\ref{mRrel}). After simplification, we arrive at
\begin{equation} \label{RDFfin}
\calf_R = 3\pi \lambda \rho R^2 \calf_m\left(\pi \lambda \rho R^3\right),
\end{equation}
where $\calf_m\left(\pi \lambda \rho R^3\right)$ implies that the function $\calf_m$ is evaluated at $m = \pi \lambda \rho R^3$. For the sake of completeness, we shall present the two limits for the width distribution below. We begin with
\begin{equation} \label{R0DF}
\lim_{R \rightarrow 0}\, \calf_R  \propto R^{-2},
\end{equation}
and it is important to reiterate that our analysis is for the \emph{3D case}. A similar result was also proposed in the 2D case by \citet{ULS10} via their phenomenological arguments and numerical results. We also refer the reader to the 2D simulations by \citet{SLMR13} and \citet{SGP16} where the same scaling law was obtained. The slopes (for the width distribution) can be evaluated in the 3D case when compared to the 2D case given the \emph{same} mass distribution. Of course, one expects the 2D and 3D cases to have different mass distributions as well, but it is still instructive to consider the above argument. Let us assume that $\calf_m \propto m^{-\Gamma}$ in the limit $m \rightarrow 0$, where $\alpha$ is positive. Then, in the 2D case, we find that
\begin{equation} \label{R2DDF}
\calf_R^{(2D)} \propto R^{-2\Gamma + 1},
\end{equation}
since $m \propto R^2$ in the 2D case. For the 3D case, we use $m \propto R^3$ that leads us to
\begin{equation} \label{R3DDF}
\calf_R^{(3D)} \propto R^{-3\Gamma + 2}.
\end{equation}
From (\ref{R2DDF}) and (\ref{R3DDF}), we see that the 3D width distribution is steeper than its 2D counterpart only when $\Gamma > 1$ is valid. It is clear from (\ref{m0DF}) that this condition is indeed met since $\Gamma = 4/3$, leading to a steeper 2D distribution. A consequence of our analysis is that the width distribution of $-2$ would not be possible in 2D, if the mass distribution were still the same as (\ref{m0DF}). The other limit for the width distribution is found from (\ref{mInfDF}), and is of the form
\begin{equation} \label{RDFInfty}
\lim_{R \rightarrow \infty}\, \calf_R  \propto R^{-5/2} e^{-\pi \gamma \lambda \rho R^3}.
\end{equation}
We have plotted (\ref{RDFfin}) in Fig. \ref{fig:distribution_R}, and the power-law limit for small $R$, namely (\ref{R0DF}), is also clearly illustrated. In a manner similar to the mass distribution, the exponential cutoff dominates when larger values of $R$ are considered.

We conclude this section by observing that a pure power law distribution for (\ref{mDF}) follows if only the energy constraint is retained, i.e. if we let $\{\gamma,\delta,\varepsilon\} = 0$. For this simplified case, we find that the $-4/3$ power law holds true for all values of $m$. Hence, the result coincides with the more general distribution function in the limit $m \rightarrow 0$ as seen from (\ref{m0DF}).

\subsection{The flux and helicity distributions}\label{SSecFHDF}
In order to obtain the total flux distribution function, we must use (\ref{mpsiDFfinal}), thereby obtaining
\begin{equation} \label{psiDFdefin}
\calf_\psi = \int_{m = 0}^{m = \infty} \calf_{m,\psi}\,dm.
\end{equation}
The integral can be performed analytically, but the result is tremendously complex since (\ref{mpsiDFfinal}) has a fairly complicated dependence on $m$. Hence, we shall not present it here, but we have duly carried out the series expansion for $\psi \rightarrow 0$. The leading order contribution is
\begin{equation} \label{psi0DF}
\lim_{\psi \rightarrow 0}\, \calf_\psi \propto \psi^{-3},
\end{equation}
indicating that the slope of the distribution function is steeper than the 2D analyses conducted by \citet{ULS10} and \citet{HB12}, who arrived at slopes of $-2$ and $-1$ respectively. But, we wish to point out two notable differences: (i) our work assumes a 3D geometry, and (ii) the flux $\psi$ that we have used is the total flux and is therefore not the same variable as the one that they had used. The fact that the slope for the 3D case is steeper than its 2D counterpart is quite consistent with the general observations for the width distribution that were presented in Sec. \ref{SSecMWDF}.

There is also another means of arriving at (\ref{psi0DF}), which we describe below. If we introduce $y = m^{-1/2}$, we have
\begin{eqnarray} \label{psiDFDef}
\calf_\psi &=& 2\left(\frac{2\pi}{\beta}\right)^{3/2} \exp\left[-\frac{\psi\left(2\delta + \varepsilon\psi\right)}{2}-1\right] \\
&& \times \int_{y=0}^{y=\infty} \exp\left[-\frac{\beta \calc}{2} \rho^{1/3} \psi^2 y^{2/3} - \gamma y^{-2}\right]\,dy. \nonumber 
\end{eqnarray}
If we let $\psi \rightarrow 0$ in the above expression, by dropping the first term, it is evident that the integral does not converge. The transformation $\tilde{y} = \psi\, y^{1/3}$ yields the factor of $\psi^{-3}$ that can be pulled outside the integral, and the latter converges to a finite value upon setting $\psi = 0$. Consequently, we have illustrated how the limit (\ref{psi0DF}) can be obtained.

In Sec. \ref{SSecMWDF}, we proved that the inclusion of the energy constraint alone or the general case led to the same result for $\calf_m$ in the limit $m \rightarrow 0$. Hence, it is instructive to study the two cases for $\psi \rightarrow 0$ as well. The general case leads to (\ref{psi0DF}), whilst the case with only the energy constraint is easily evaluated by setting $\{\gamma,\delta,\varepsilon\} = 0$. The integral can be evaluated analytically by using the fact that $\int_{r=0}^{r=\infty} \exp\left(-A r^{2/3}\right)\,dr \propto A^{-3/2}$. We end up with (\ref{psi0DF}), as before, thereby confirming that the two cases yield identical results when $\psi \rightarrow 0$.

Let us now proceed to evaluate the other limit. We shall formally consider the case where the coefficients preceding the two terms inside the integral (\ref{psiDFDef}) are very ``large''. In general, note that this condition will be fully satisfied when $\psi \rightarrow \infty$. With this assumption, the first term in the expression is unity for small $y$ and rapidly falls off at large $y$, whilst the opposite behavior is seen for the second term. 

This enables us to construct a heuristic approximation along the lines described in \citet{DB58,BH75,Hin91} for Laplace's method \footnote{Alternatively, we could introduce the change-of-variables $\bar{y}=\psi^{3/4}\,{y}$, thereby enabling us to directly apply the conventional form of Laplace's method. The asymptotic expansion of the integral for $\psi \rightarrow \infty$ follows as a natural consequence.}. The trick is to Taylor expand the function (inside the exponential) around the maximum, and carry out the integration. Denoting the integral by $I$, we find that it simplifies to
\begin{equation} \label{SPInt}
I \approx \exp\left(-\frac{4\gamma}{y_\star^{2}}\right) \int_{y=0}^{y=\infty} \exp\left[-\frac{8\gamma}{3} y_\star^{-4} \left(y - y_\star\right)^2\right]\,dy,
\end{equation}
where we have introduced the variable
\begin{equation}
y_\star = \left(\frac{\beta \calc \rho^{1/3} \psi^2}{6\gamma}\right)^{-3/8}.
\end{equation}
(\ref{SPInt}) can now be further simplified to yield
\begin{equation} \label{Iinfdef}
I \approx \sqrt{\frac{3\pi}{8\gamma}}\, y_\star^2 \exp\left(-\frac{4\gamma}{y_\star^{2}}\right),
\end{equation}
where we have used the fact that $\psi \rightarrow \infty$ to eliminate an error function $\left(\mathrm{Erf}\right)$ that approaches unity. Hence, we can conclude that
\begin{equation}
\lim_{\psi \rightarrow \infty}\, \calf_\psi \propto \psi^{-3/2} \exp\left(-\frac{\varepsilon\psi^2}{2}\right),
\end{equation}
where we have chosen to deliberately retain only the highest power of $\psi$ in the exponential factor arising from (\ref{psiDFDef}) and (\ref{Iinfdef}). In general, there are also terms involving $\psi$ and $\psi^{3/2}$ inside the exponential function. 

\begin{figure}
  \centering
  \includegraphics[width=1.0\linewidth]{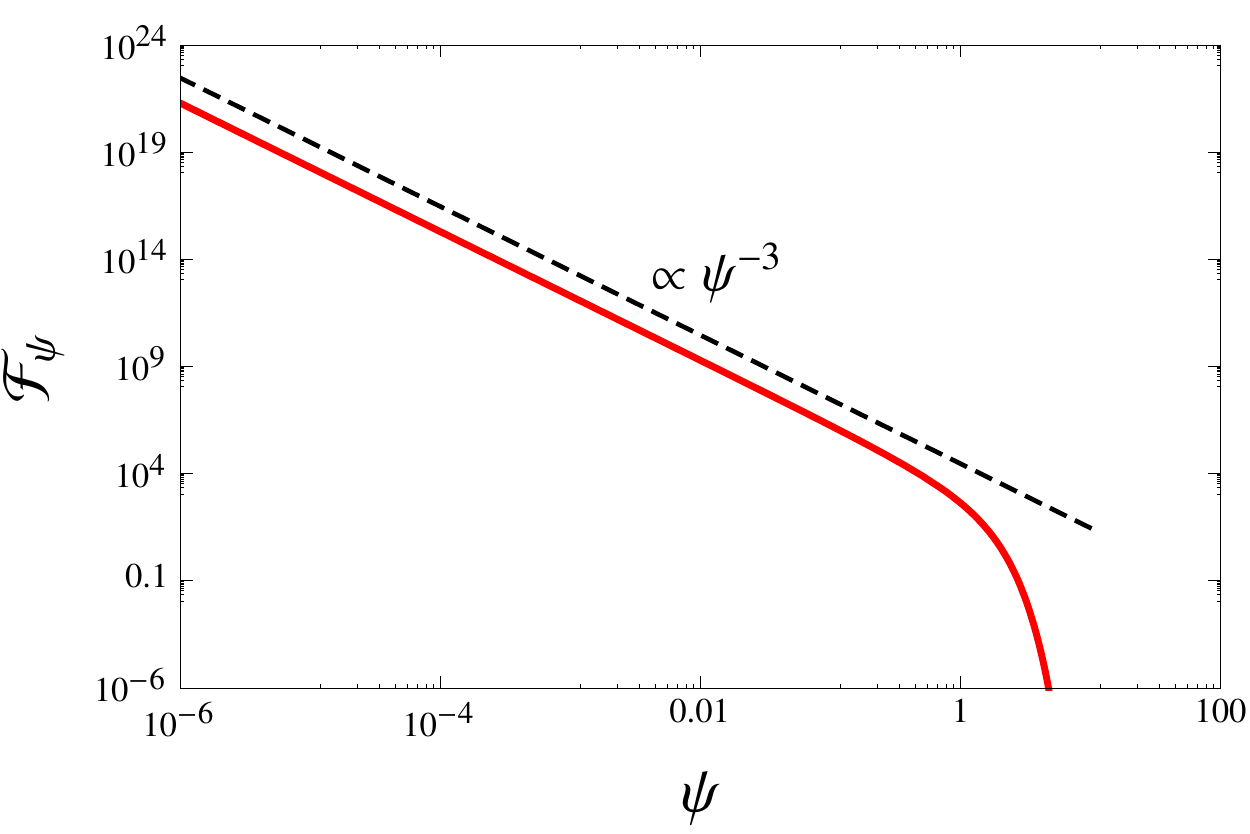}
  \caption{The total flux distribution function (\ref{psiDFdefin}) is plotted as a function of $\psi$. The solid red curve represents the numerical solution of (\ref{psiDFDef}). The dashed black line denotes the asymptotic solution (\ref{psi0DF}), which works well for small $\psi$. Here, we have used the fiducial values $\beta = \gamma = \delta = \varepsilon = 1$ and $\calc = 1/4\pi$. The latter value was chosen because this factor appears in the magnetic energy.}
  \label{fig:distribution_psi}
\end{figure}

We have plotted the numerical solution of (\ref{psiDFDef}) in Fig. \ref{fig:distribution_psi}. Akin to the mass (or width) distribution, the existence of an exponential falloff at large values of $\psi$ is quite apparent. The veracity of the power law analytical limit (\ref{psi0DF}) for small $\psi$ has also been verified through comparison with the numerical solution of (\ref{psiDFDef}). 

In Sec. \ref{SSecCons}, it was pointed out several times that the total flux and the helicity are proportional to one another. The constant of proportionality is taken in some instances to be $0.5$ \citep{BHPvS}, and this was also adopted in (\ref{Helicity}). We use the property of probability conservation in a differential area to write
\begin{equation}
\calf_\psi d\psi = \calf_\calh d\calh,    
\end{equation}
where $\calh = \psi^2/2$ and $\calf_\calh$ denotes the helicity distribution of plasmoids. Thus, we find that
\begin{equation} \label{HDFdef}
\calf_\calh = \sqrt{\frac{2}{\calh}} \calf_\psi\left(\sqrt{2\calh}\right),    
\end{equation}
and $\calf_\psi$ is evaluated at $\psi = \sqrt{2\calh}$. The two limiting cases of $\calf_\calh$ are therefore found to be
\begin{equation} \label{Hel0DF}
 \lim_{\calh \rightarrow 0}\, \calf_\calh \propto \calh^{-2},
\end{equation}
\begin{equation}
 \lim_{\calh \rightarrow \infty}\, \calf_\calh \propto \calh^{-5/4} \exp\left(-\varepsilon \calh\right),
\end{equation}
and we have retained only the highest power in the exponential factor in the above expression.

\begin{figure}
  \centering
  \includegraphics[width=1.0\linewidth]{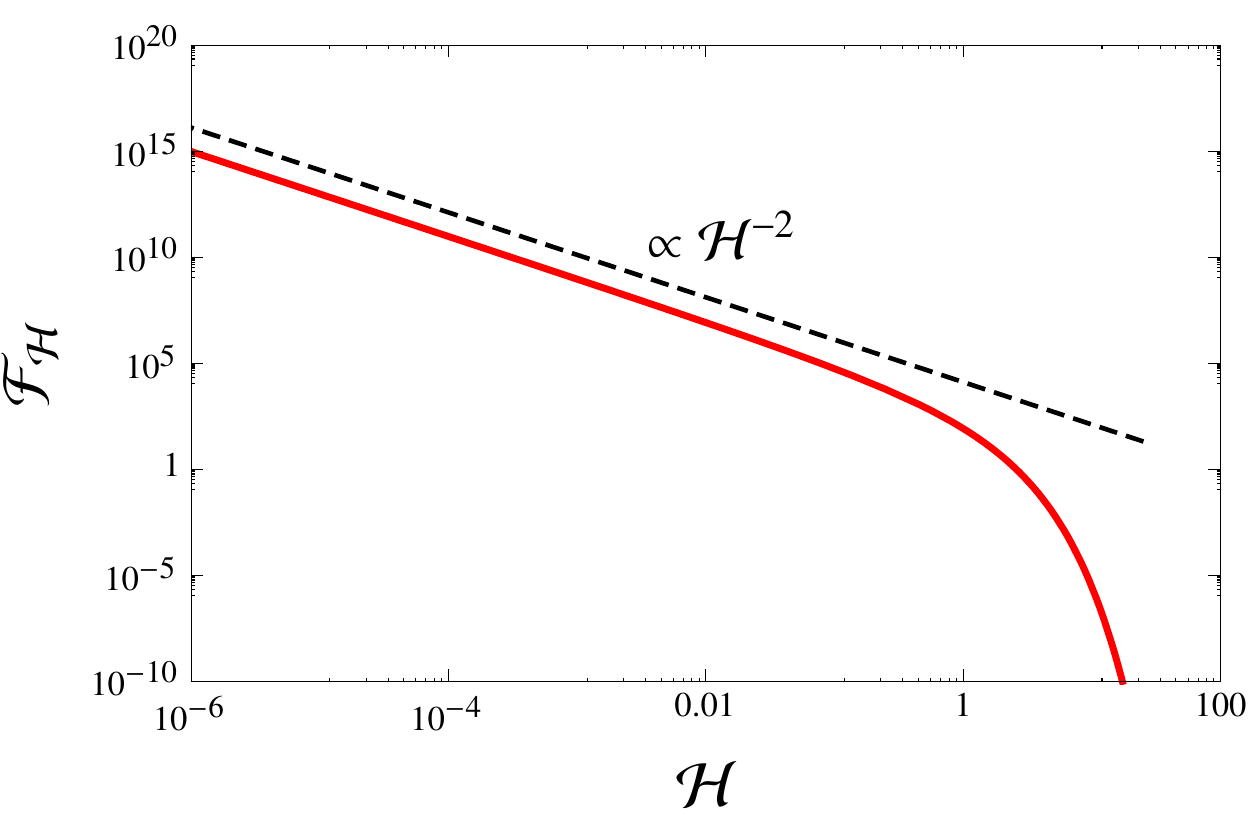}
  \caption{The helicity distribution function is plotted as a function of the helicity $\calh$. The solid red curve represents the numerical solution of (\ref{HDFdef}). The dashed black line denotes the asymptotic solution (\ref{Hel0DF}), which is valid for small $\calh$. Here, we have used the fiducial values $\beta = \gamma = \delta = \varepsilon = 1$ and $\calc = 1/4\pi$. The latter value was chosen because this factor appears in the magnetic energy.}
  \label{fig:distribution_H}
\end{figure}

We have plotted the numerical solution of the helicity distribution function in Fig. \ref{fig:distribution_H}. This has been obtained by using the numerical solution for $\calf_\psi$ in conjunction with (\ref{HDFdef}). The limiting behavior for small $\calh$, which is seen to be a power law from the figure, also matches the analytical limit obtained in (\ref{Hel0DF}). For higher values of $\calh$, the exponential cutoff is increasingly dominant.
 
\subsection{The velocity distribution} \label{SSecVelDF}
To compute the velocity distribution, we first introduce
\begin{equation} \label{vmDFdefn}
    \calf_{m,{\bf v}} = \int_{\psi=0}^{\psi=\infty} \calf\,d\psi,
\end{equation}
which leads to the analytical expression
\begin{eqnarray} \label{mvDFexp}
\calf_{m,{\bf v}} &=& \sqrt{\frac{\pi}{2\sigma}} \left[1-\mathrm{Erf}\left(\frac{\delta}{\sqrt{2\sigma}}\right)\right] \nonumber \\
&& \quad \times \exp\left(\frac{\delta^2}{2\sigma}-\frac{\beta m v^2}{2}-\gamma m - 1\right),
\end{eqnarray}
where $\sigma$ was defined in (\ref{defsigma}). The velocity distribution follows from
\begin{equation} \label{DFvDef}
\calf_{\bf v} = \int_{m=0}^{m=\infty} \calf_{m,{\bf v}}\,dm.
\end{equation}
The integral cannot be performed analytically, and even the asymptotic expansions are somewhat intricate. Hence, we shall resort to an alternate strategy to deduce the two limits. From (\ref{defsigma}), we see that $0 < \sigma^{-1} < 1/\varepsilon$, and the two limits are attained at $m = 0$ and $m = \infty$ respectively. This implies that
\begin{equation}
1-\mathrm{Erf}\left(\frac{\delta}{\sqrt{2\varepsilon}}\right)\ < 1-\mathrm{Erf}\left(\frac{\delta}{\sqrt{2\sigma}}\right) < 1,
\end{equation}
and, similarly, we also find
\begin{equation}
1 < \exp\left(\frac{\delta^2}{2\sigma} \right) < \exp\left(\frac{\delta^2}{2\varepsilon} \right).
\end{equation}
The above relations indicate that the two functions are bounded by finite values from above and below and their product is also expected to have finite lower and upper bounds. Hence, we do not expect the functional dependence of (\ref{DFvDef}) on ${\bf v}$ to be significantly altered by these functions, especially for certain choices of $\delta$ and $\varepsilon$. In contrast, $\sigma^{-1}$ and $\exp\left(-\zeta m\right)$ do not have finite lower bounds since they vanish for $m \rightarrow 0$ and $m \rightarrow \infty$ respectively, where
\begin{equation} \label{zetadef}
\zeta = \frac{\beta v^2}{2} + \gamma.
\end{equation}

\begin{figure}
  \centering
  \includegraphics[width=1.0\linewidth]{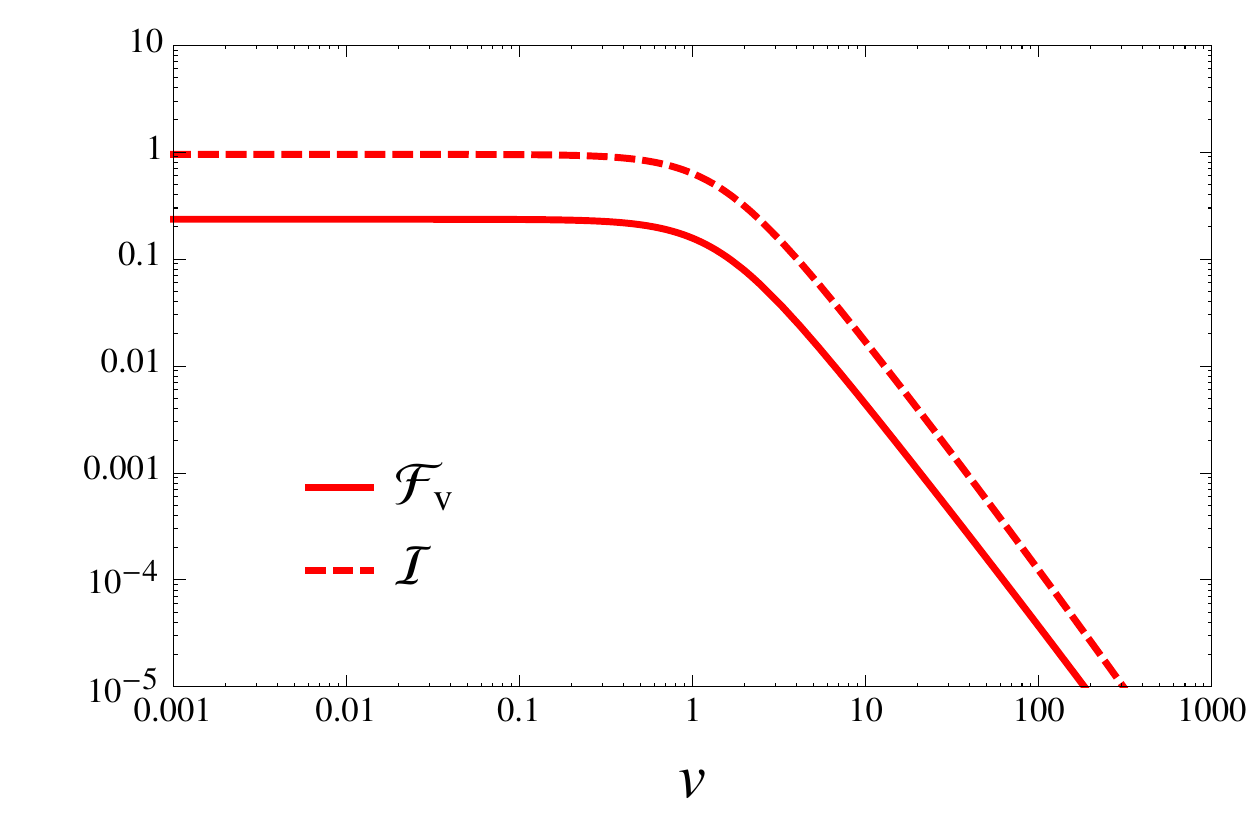}
  \caption{The solid curve represents $\calf_{\bf v}$, and its numerical solution is computed from (\ref{DFvDef}). The dashed curve signifies the exact analytical expression for the integral $\cali$, given by (\ref{caliDef}). For both the curves, we have used the fiducial values $\beta = \gamma = \delta = \varepsilon = 1$ and $\calc = 1/4\pi$. The latter value was chosen because this factor appears in the magnetic energy.}
  \label{fig:distribution_FvsI}
\end{figure}

Hence, we shall center our attention on the integral
\begin{equation} \label{MeiInt}
\cali = \int_{m=0}^{m=\infty} \frac{\exp\left(-\zeta m\right)}{\sqrt{\varepsilon + \beta \calc \rho^{1/3} m^{-1/3}}}\,dm,
\end{equation}
which will help us elucidate the general behavior of $\calf_{\bf v}$. Our approach is motivated partly by classical asymptotic analysis \citep{DB58,BH75,BenOr78}, but it has a greater range of validity since there are no small or large parameters involved. We have verified that the full numerical solution of (\ref{DFvDef}) is well represented by the integral (\ref{MeiInt}), which has the analytical solution (\ref{caliDef}) discussed below. A comparison of these two plots is provided in Fig. \ref{fig:distribution_FvsI}, and the accuracy of (\ref{MeiInt}) stands out. Observe that the two curves are not exactly the same, since we had neglected some constant numerical factors such as $e^{-1}$ whilst defining (\ref{MeiInt}).

After a careful evaluation of the integral, the result is analytical, expressible in terms of the Meijer G-function \citep{BaEr53,AbSte65}, given by
\begin{equation} \label{caliDef}
\cali = \frac{\sqrt{3}\,\beta^3\calc^3 \rho}{4\pi^{5/2}\,\varepsilon^{7/2}}\,\,\MeijerG[\Bigg]{4}{3}{3}{4}{-\tfrac{5}{6},\, -\tfrac{1}{2},\, -\tfrac{1}{6}}{-1,\, -\tfrac{2}{3},\, -\tfrac{1}{3},\, 0}{\frac{\beta^3\calc^3 \rho \zeta}{\varepsilon^3}}.
\end{equation}
If we take the limit ${\bf v} \rightarrow 0$, it follows that $\zeta \rightarrow \gamma$. This leads us to
\begin{equation} \label{DFvZero}
 \lim_{{\bf v} \rightarrow 0}\, \calf_{\bf v} \propto \left(v^2\right)^0.
\end{equation}
In the other case, we consider ${\bf v} \rightarrow \infty$, which is equivalent to letting $\zeta \propto v^2$ to leading order, evident from the definition (\ref{zetadef}). The asymptotic expansion of (\ref{caliDef}) for large values of $\zeta$ enables us to conclude that $\cali \propto \zeta^{-7/6}$. Rewriting this expression in terms of $v$, we end up with 
\begin{equation} \label{DFvInfinity}
 \lim_{{\bf v} \rightarrow \infty}\, \calf_{\bf v} \propto \left(v^2\right)^{-7/6}.
\end{equation}
Hence, we have successfully arrived at the two asymptotic limits for $\calf_{\bf v}$, and they are both power laws. Such behavior is rather unusual since the other distribution functions exhibited an exponential falloff for larger values of their respective variables. 

Suppose that we consider the case where all the constraints apart from the energy are dropped, which amounts to setting $\{\gamma,\delta,\varepsilon\} = 0$ in (\ref{mvDFexp}).  The integral (\ref{DFvDef}) can then be performed analytically via the use of the Gamma function, and a power law behavior in $v^2$ is observed. The exponent turns out to be exactly the same as the one obtained in  (\ref{DFvInfinity}). In this regard as well, the velocity distribution function is rather unusual. For the mass and flux distribution functions, retaining only the energy gave the same power-law behavior for \emph{small} values. In contrast, we have seen that the same power-law behavior for  the velocity distribution is obtained for \emph{large} values of ${\bf v}$. 

Finally, a clarification regarding $\calf_{\bf v}$ is necessary. Although we have determined, and plotted, it as a function of $v$, the distribution function $\calf_{\bf v}$ is really a function of ${\bf v}$. This stems from the fact that it actually depends on $v^2 \equiv {\bf v} \cdot {\bf v}$. In other words, the distribution function is isotropic in ${\bf v}$, but it encompasses the full 3D velocity dependence. If our system was endowed with momentum or angular momentum dependence, the resultant distribution function would have been anisotropic - a similar situation also exists in gravitational dynamics \citep{BT87}. 

\begin{figure}
  \centering
  \includegraphics[width=1.0\linewidth]{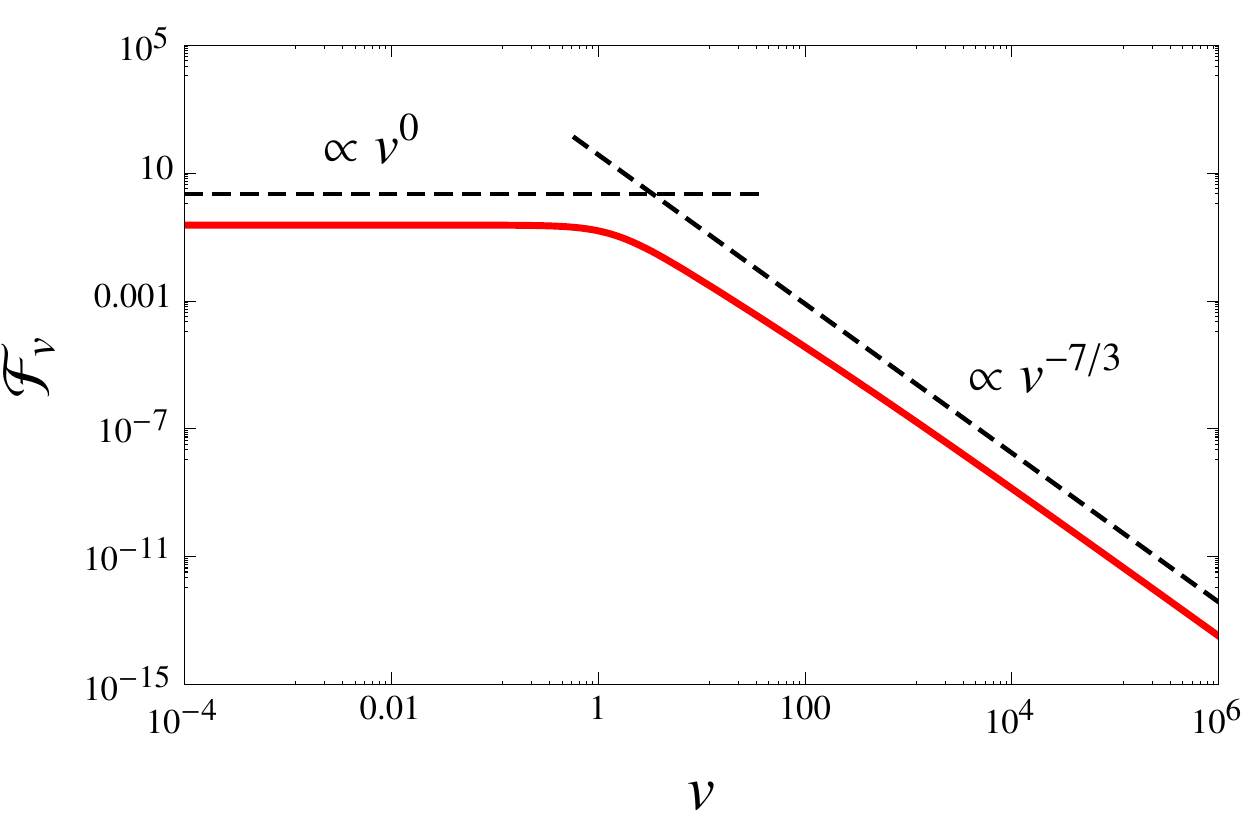}
  \caption{The velocity distribution function is plotted as a function of the modulus of the total velocity. The solid red curve represents the numerical solution of (\ref{DFvDef}). The dashed and dotted-dashed black lines denote the asymptotes (\ref{DFvZero}) and (\ref{DFvInfinity}), which are valid for small and large values of ${\bf v}$ respectively. Here, we have used the fiducial values $\beta = \gamma = \delta = \varepsilon = 1$ and $\calc = 1/4\pi$. The latter value was chosen because this factor appears in the magnetic energy.}
  \label{fig:distribution_v}
\end{figure}

In Fig. \ref{fig:distribution_v}, the numerical solution of the 3D velocity distribution function (\ref{DFvDef}) is provided. Note that we have plotted $\calf_{\bf v}$ as a function of $v$ (instead of $v^2$) since this variable is more physically transparent for visualization purposes. An inspection of the figure reveals that the analytical asymptotes (\ref{DFvZero}) and (\ref{DFvInfinity}), valid for small and large values of ${\bf v}$ respectively, capture the correct behavior of $\calf_{\bf v}$ in these regimes. Unlike all of the previous distribution functions, there is no exponential cutoff since we observe a power-law behavior in both limits.

One can also compute the distribution function for the modulus of the velocity alone, viz. effectively a 1D distribution, denoted by $\calf_v$ hereafter. Using the probability conservation law described in Sec. \ref{SSecMWDF}, we have
\begin{equation}
\calf_{\bf v} d{\bf v} = \calf_v dv,
\end{equation}
and we can now use $d{\bf v} = 4\pi\,v^2\,dv$ as per the discussion in Sec. \ref{SSecMWDF}. Note that $\calf_{\bf v}$ must be expressed in terms of $v$, which has already been done since $\calf_{\bf v}$ depended on $v^2$. Hence, we end up with
\begin{equation}
\calf_v = 4\pi v^2\,\calf_{\bf v},
\end{equation}
and the LHS and RHS are solely functions of $v$. We are now in a position to derive the two limiting cases of $F_v$, which are presented below.
\begin{equation}
 \lim_{v \rightarrow 0}\, \calf_v \propto v^2,
\end{equation}
\begin{equation}
 \lim_{v \rightarrow \infty}\, \calf_v \propto v^{-1/3},
\end{equation}
and this implies that the number of particles increases with $v$, attains a maximum, and then decreases for higher values of $v$. 

\section{The connections with space and astrophysical plasmas} \label{SecAstroImp}
Since the 1970s and 1980s, detailed observations of flux ropes/tubes (that we have referred to as ``plasmoids'') in the Earth's magnetotail have been available \citep{RE78,PHPS82,SRS84}. It was also during this period that a great deal of progress, both from a theoretical and observational standpoint, was achieved concerning the formation and properties of plasmoids \citep{Hon77,Hon79}. By the 1980s, it was well understood that plasmoids were inherently three-dimensional in nature, corresponding to flux rope/tube structures \citep{Hon82,HuSi87}. In contrast, many theoretical and numerical studies carry out 2D analyses. 

There have been many statistical analyses of 3D plasmoids (flux ropes) over the past few decades in the terrestrial, solar and planetary contexts \citep{Rich87,MH92,Ichet98,HoSo01,Viget04,Jack11,Vogt14}. From these studies, a few broad inferences can be drawn, and we shall begin by considering the data from the magnetotail.
\begin{itemize}
\item It has been shown that the width distribution is subject to an exponential falloff for large values of $R$ (Fig. 7 of \citet{FDSH11}), which is loosely consistent with (\ref{RDFInfty}). 
\item Earlier papers by \citet{Rich87} (see Fig. 16) and \citet{MH92} (see Fig. 13) also appear to show a somewhat similar trend.
\item In Fig. 9 of \citet{MH92}, the velocity distribution declines for larger $v$, thereby proving to be qualitatively similar to the trend predicted in Sec. \ref{SSecVelDF}.
\end{itemize}
Furthermore, in the Martian magnetosphere, the width distribution may also display an exponential falloff for large values, as predicted by (\ref{RDFInfty}) - see Fig. 4b of \citet{BriBra11}. However, we wish to emphasize that the above statements must be viewed with due caution for the following reasons: (i) the number of plasmoids observed is typically low, (ii) the interactions between the plasmoids are likely to be relatively insignificant, and (iii) resistive MHD is not applicable to these systems. Hence, it is more instructive to consider the results from solar observations on account of the reasons that were identified in Secs. \ref{SecIntro} and \ref{SecMaxEnt}.

There exist some commonalities between the flux rope structures observed in the magnetotail and in coronal mass ejections (CMEs), as noted in \citet{LCF08} and \citet{LM09}. In the latter context, several statistical analyses of solar flux rope properties (such as the width) have been conducted in recent times. In \citet{JDD14}, small flux ropes \citep{MoFo00} were shown to obey a power-law width distribution with an exponent of $-2.4$ (other cases such as $-1.8$ and $-2.1$ were also considered), whilst magnetic clouds (larger flux ropes) were characterized by a Gaussian distribution. 

These results are mostly consistent with our theoretical predictions (\ref{R0DF}) and (\ref{RDFInfty}), especially the former where the power-law exponent of $-2$ was derived. We also refer the reader to \citet{FWC07} and \citet{Feng08} for the size distributions of solar flux ropes, which were also analyzed by \citet{JDD14} in their work. A summary of the current status of these statistical studies can be found in Section 4.3 (see also Fig. 17) of \citet{Jan16}. The combination of a power law at small scales, and the exponential cutoff at larger scales was also obtained in \citet{GBH13} for post-CME sheets through the use of topological methods. Clearly, this result is qualitatively consistent with our methodology which predicts the absence of a universal power law across all scales. 

Although \citet{GBH13} suggest that the power-law exponent for small values may be approximately $-1$, this prediction is partially motivated by the relation $\psi \propto R$ that was proposed in \citet{ULS10}. Moreover, our result does not appear to directly contradict their findings since the limit (\ref{R0DF}) is formally valid only in the limit $R \rightarrow 0$. The analytical prediction in this regime cannot be easily probed by either simulations or data analysis (since there are an insufficient number of plasmoids).

Hitherto, we have focused on comparing the general trends (and specific results) predicted by our model with some of the existent observations. We close this Section by pointing out a couple of avenues where our scalings may prove to be useful in understanding astrophysical systems. The first is the capacity of plasmoid motions and plasmoid-mediated reconnection for driving particle acceleration, which has been studied extensively \citep[e.g][]{Oi02,DSCS06,Chet08,Oka10,Birn12,LCP13,NiShi13,NKJB15,NUCWB15,IGHB15}. The velocity distribution of the plasmoids computed herein may therefore enable a better understanding of how particle acceleration can occur via interactions with plasmoids.

Lastly, we observe that plasmoid reconnection has been studied extensively in the context of solar phenomena such as flares and CMEs \citep{KKB00,ShiTa01,Bart2011,SIH13,LMS15,ShiTa16,Dud16}. Most of these processes are likely to be operational on other stars as well \citep{MSN12,Shi13,Lin17}. A better understanding of the statistical properties of the ejected plasmoids could contribute to our understanding of the distribution of superflares and CMEs on M-dwarfs \citep{CHMBS14,ShiTa16}. In turn, we expect that gaining such knowledge would enable us to make some progress towards understanding the prebiotic chemistry \citep{SWMKH,Aira16,Ling17} and atmosphere escape rates of exoplanets, especially around M-dwarfs \citep{SBJ16,DL17,LL17}.

\section{Conclusion} \label{SecConc}
In statistical physics, the Principle of Maximum Entropy has proven to be very successful in deducing the probability distribution function given a set of invariants. The advantage of this approach has been argued to lie in its simplicity, physical transparency and versatility; the latter arises from the fact that this method can be interpreted as a form of statistical inference that can be applied to near-equilibrium systems \citep{Jay03}.

After adopting this approach, we arrived at the final distribution function for 3D plasmoids, which depended on the mass, velocity and total flux, each of which constitutes physically meaningful variables. Our treatment is primarily applicable to systems with large numbers of interacting plasmoids, and where resistive MHD constitutes a reasonable description of the underlying dynamics. To the best of our knowledge, a 3D analysis along these lines is novel, and the distributions for these variables have never been derived before. Our final expression for the three-variable distribution function, viz. (\ref{EqDF}), depended on four free parameters, as well as another one that quantified the geometric shape of the plasmoids. At first glimpse, one may therefore argue that there is a high degree of arbitrariness, owing to the large number of unspecified parameters.  

However, we showed that a certain degree of generality (within the resistive MHD model) still emerged when the mass, width, total flux, and helicity distribution functions were determined. They all possessed a power-law behavior for small values, and were subject to an exponential falloff for large values. Moreover, the power-law exponents at small values were very robust, and were nearly independent of the choices of free parameters. The power-law exponents were $-4/3$, $-2$, $-3$ and $-2$ for the mass, width, flux, and helicity distributions respectively. Even for larger values, the exponential functions were shown to have some commonality, regardless of the actual choices of the parameters.

The emergence of this behavior was even more striking when the distribution function for the (3D) velocity was considered. We showed that the small and large limits of the velocity distribution \emph{both} led to power laws, with exponents of $0$ and $-7/3$, when computed as a function of the variable $v = |{\bf v}|$. The two exponents remained mostly unaffected by the choices of the free parameters prevalent in our overall distribution function, which lends credence to the hypothesis that these asymptotes may be quite robust insofar our model is concerned.

Finally, we have undertaken a preliminary comparison of our model with the observational data, and concluded that the former yields results that are mostly consistent with the latter. It must, however, be recalled that this study was somewhat rudimentary, and in-depth observational (and numerical) data analyses \citep{SRW16,NZML17} are undoubtedly necessary for obtaining a clear picture of the strengths and weaknesses of the theoretical model. Along the way, we also delved briefly into the attendant astrophysical implications for phenomena ranging from particle acceleration to solar flares and CMEs.

As the formalism proposed in this paper might be applicable to other scenarios, several avenues open up for subsequent research. The same analysis can be redone by introducing a new set of constraints for the 2D system that are clearly distinct from the 3D case \citep{kraichnan80,MH84}, and the ensuing scaling laws are also expected to be different compared to their 3D versions. 

Another possibility is to obtain the distribution of relativistic plasmoids, which could then be compared against recent simulations \citep{SGP16,PCS17}. The inclusion of turbulent kinetic and magnetic energies might also pave the way for analyzing the statistical properties of turbulent plasmoid-mediated reconnection \citep{HuBha16}; see also \citet{LEVK15}. Based on the simplicity and elegance of the MaxEnt approach, we suggest that further theoretical analyses along these lines may be warranted. 

\acknowledgments
The authors are very grateful to Amitava Bhattacharjee and Yi-Min Huang for the enlightening discussions and the perceptive remarks. The beneficial conversations with Eero Hirvijoki, Russell M. Kulsrud, Takuya Shibayama and Roscoe B. White are also acknowledged. The authors were supported by the Department of Energy Contract No. DE-AC02-09CH11466 and the National Science Foundation Grant Nos. AGS-1338944 and AGS-1552142 during the course of this work.

%\bibliographystyle{apsrev4-1}
%\bibliography{DistPlasm}


\end{document}